*Research Article*

# The Protective Effects of the Ethyl Acetate Part of Er Miao San on Adjuvant Arthritis Rats by Regulating the Function of Bone Marrow-Derived Dendritic Cells

**Jiemin Ding,[1,2,3] Min Liu,[1,2,3] Zihua Xuan,[1,2,3] Meng li Liu,[1,2,3] Ning Wang,[1,2,3] and Xiaoyi Jia[1,2,3]**

[1]*School of Pharmacy, Anhui University of Chinese Medicine, Hefei 230012, China*
[2]*Anhui Province Key Laboratory of Chinese Medicinal Formula, Hefei, Anhui 230012, China*
[3]*Anhui Province Key Laboratory of Research & Development of Chinese Medicine, Hefei 230012, China*

Correspondence should be addressed to Ning Wang; wnsci123@163.com and Xiaoyi Jia; jiaxiaoyi78@126.com

*Aims*. The aim of this study was to evaluate the protective effects of Er Miao San (EMS) and the regulative function of bone marrow-derived dendritic cells (BMDCs) on adjuvant arthritis (AA) in rats. *Methods*. The ethyl acetate part of EMS (3 g/kg, 1.5 g/kg, and 0.75 g/kg) was orally administered from day 15 after immunization to day 29. The polyarthritis index and paw swelling were measured, the ankle joint pathological changes were observed using hematoxylin-eosin (HE) staining, and the spleen and thymus index were determined. Moreover, T and B cell proliferation were determined using the CCK-8 assay. The expression of BMDC surface costimulatory molecules and inflammatory factors were determined using flow cytometry and ELISA kits, respectively. *Results*. Compared with the AA model rats, the ethyl acetate fraction of EMS obviously reduced paw swelling (from 1.0 to 0.7) and the polyarthritis index (from 12 to 9) ($P < 0.01$) and improved the severity of histopathology ($P < 0.01$). The treatment using ethyl acetate fraction of EMS significantly reduced the spleen and thymus index ($P < 0.01$) and inhibited T and B cell proliferation ($P < 0.01$). Moreover, EMS significantly modulated the expression of surface costimulatory molecules in BMDCs, including CD40, CD80, CD86, and major histocompatibility complex class II (MHC-II) ($P < 0.01$). The results also showed that the ethyl acetate part of EMS significant inhibited the levels of proinflammatory cytokines interleukin- (IL-) 23 tumor necrosis factor- (TNF-) $\alpha$ and inflammatory factor prostaglandin (PG) E2 in the supernatant of BMDCs. However, the level of anti-inflammatory cytokine IL-10 was significantly increased ($P < 0.01$). *Conclusion*. These results suggest that the ethyl acetate part of EMS has better protective effects on AA rats, probably by regulating the function of BMDCs and modulating the balance of cytokines.

## 1. Introduction

Rheumatoid arthritis (RA) is an autoimmune disease characterized by joint synovial inflammation and cartilage damage [1]. An abnormal proliferation of fibroblast synovial cells is a typical feature of RA, which is correlated with over-activated immune cells, such as T-cells, B-cells, macrophages, and dendritic cells (DCs). A large number of infiltrated DCs are highly involved in the synovium of RA patients and arise to the occurrence of arthritic disease.

DCs are specialized antigen-presenting cells (APCs), which can take up processes, present antigens, and initiate T-cell-mediated immune responses [2]. DCs highly express surface costimulatory molecules, including CD40, CD80, CD86, and MHC-II, which present antigens to T-cells. Mature DCs can effectively activate the initial T cells, which are the center of initiation, regulation, and maintenance of immune responses [3]. Previous studies have confirmed the incidence and development of RA related to DCs. The synovial DCs of RA patients secrete chemokines and attract proinflammatory immune cells, including macrophages and neutrophil monocytes [4, 5]. Compared with healthy controls, the concentration of proinflammatory cytokines (IL-1$\beta$, IL-6, and IL-23) and inflammatory factors (PGE2) in RA patients have increased [6, 7]. This is one of the key factors of RA pathogenesis.

Er Miao San (EMS) is a traditional Chinese Medicine prescription, which is derived from《dan xi xin fa》. It consists of Phellodendri Cortex and Atractylodis Rhizoma. We have previously reported that the ethyl acetate part of EMS could significantly reduce the pathological changes, having a therapeutic effect on adjuvant-induced arthritis (AA) rats [8]. Nevertheless, the immune mechanisms targeted by the ethyl acetate part of EMS are unclear. In this research, the aim was to investigate the antiarthritis mechanisms of EMS, whether its potential molecular mechanism was related to the regulation of the function of DCs.

## 2. Materials and Methods

*2.1. Animals.* Sprague Dawley rats (male, $180 \pm 20$ g) were purchased from the Animal Department of Anhui Medical University, China. Rats were adapted to standard laboratory conditions (under a controlled temperature of 22–26°C and a 12 h light and 12 h dark period) and to feed criterion forage and water daily during the experiment. All experiments were approved by the Experimental Animal Ethics Committee of Anhui University of Chinese Medicine (Identification number: 202005).

*2.2. Reagents.* Phellodendri Cortex and Atractylodis Rhizoma were purchased from Anhui herbal pieces Co., Ltd. (BoZou, Anhui Province, China). Petroleum ether was obtained from Shanghai SuYi Chemical Reagent Co., Ltd. Ethyl acetate was acquired from Jiangsu Qiangsheng Functional Chemical Co., Ltd. Methotrexate (MTX) was obtained from Xinyi Medical Limited Company (Shanghai, China). FITC-CD40, PE-CD80, PE-CD86, and PE-MHC-II were obtained from eBioscience, Inc. (CA, USA). Roswell Park Memorial Institute (RPMI)-1640 medium and fetal bovine serum (FBS) were obtained from Hyclone. TNF-$\alpha$, IL-10, and PGE2 were obtained from the ELISA KIT (Multi Sciences Biotech, Co., Ltd), and IL-23 was obtained from the ELISA KIT (Cusabio Biotech, Co., Ltd).

*2.3. Preparation of the Ethyl Acetate Part of EMS.* Equal parts of Atractylodis rhizoma and Phellodendri cortex were mixed and crushed. The mixture was decocted with boiling water three times for 1.5, 1.0, and 0.5 h, and the suspension was condensed by means of evaporation in a water bath at a temperature of approximately 60°C. The suspension was concentrated to 500–800 mL and extracted five times with an equivalent volume of petroleum ether. The petroleum ether part of the EMS was discarded and extracted five times with an equivalent volume of ethyl acetate. The ethyl acetate part of the EMS was concentrated to a certain concentration (0.3 g/mL, 0.15 g/mL, and 0.075 g/mL) (calculated using the crude drug).

*2.4. Induction of the AA Model and Treatment.* The AA model was constructed in SD rats, as previously described [9]. Bacillus Calmette–Guerin (BCG) (80°C, 1 h) was adequately mingled with liquid paraffin, which was complete Freund's adjuvant (CFA, 10 mg/mL). The rat AA model was injected with 0.1 mL of CFA in the left hind plantar, and the normal group was injected with 0.1 mL of an equivalent volume of saline. On day 15 after immunization, rats were randomly divided into the following groups: the normal group, AA model group, EMS (3 g/kg, 1.5 g/kg, 0.75 g/kg), and MTX (0.5 mg/kg). EMS was administered via gavage for 14 days (once per day), and the MTX group was administered via gavage every 3 days for a total of five times [10]. Meanwhile, the normal and AA model groups were orally administered an equivalent volume of a carboxymethylcellulose aqueous solution (10 mL/kg).

*2.5. Evaluation of Arthritis.* All rat weights were recorded weekly using an electronic scale. The right hind paw volume was measured using a volume meter (PV-200, Chengdu Technology Market Co., Ltd.) before immunization (basic value, day 0) and after immunization on days 14, 17, 20, 23, 26, and 29. The polyarthritis index was graded from 0 points to 16 points, as described previously: 0, no swelling normally; 1, erythema and slight swelling of the ankle joint; 2, erythema and slight swelling of the ankle joint to the metatarsophalangeal or capsular joint; 3, erythema and moderate swelling of the ankle to the metatarsophalangeal or capsular joint; and 4, erythema and severe swelling of the ankle to the metatarsophalangeal joint [11].

*2.6. Histopathological Examination of the Ankle Joints.* The rats were sacrificed on day 29 after the first immunization. The right ankle joints and the surrounding soft tissues, such as muscles and tendons, were removed. The ankle joints were obtained and fixed in a 10% formalin solution. Additionally, the ankle tissues were decalcified in 5% nitric acid before being embedded in paraffin. Paraffin sections (5 mm thick) were stained with HE. Changes in the ankles were histopathologically appraised by two blinded observers. The severity of arthritis in the ankle joints was graded from 0 to 4, according to inflammatory cell infiltration, synovial proliferation, pannus formation, and bone erosion.

*2.7. Spleen and Thymus Index.* The rats were sacrificed on day 29 after the first immunization. The food was evacuated 12 h before. Spleen and thymus tissues were removed from the body, weighed, and recorded. The spleen and thymus index were calculated as the ratio of spleen and thymus to the rat body (mg/g).

*2.8. T- and B-Lymphocyte Proliferation.* The rat spleen and thymus tissues were aseptically removed, ground using a 120 nylon cloth, spread in RPMI 1640 supplemented with 10% FBS at a density of $1 \times 106$ cells/mL, and seeded into 96-well flat-bottom plates ($1 \times 105$ cells/well). Furthermore, T- and B-cells were separately stimulated with 5 mg/L concanavalin (Con) A (Sigma) or 4 mg/L lipopolysaccharides (LPS) (Sigma) and cultured for 48 h. Two hours before the end of culture, CCK-8 (10 $\mu$L) was added to each well. The absorbance was read at 450 nm using a Microplate reader (EJ301, Thermo Fisher Scientific).

*2.9. Culture of BMDCs.* The femur and tibia were removed in a sterile environment, and the epiphysis of both ends was cut off to expose the bone marrow cavity. The bone marrow cavity was rinsed with RPMI-1640 medium and filtered using a 200 mesh nylon cloth. Cells were, then, cultured in RPMI 1640 supplemented with 10% FBS at a density of $1 \times 106$ cells/mL in normal conditions (37°C and 5% $CO_2$). The nonadherent cells were removed after 24 h of culture, and fresh culture medium, including human granulocyte-macrophage colony-stimulating factor (GM-CSF) and IL-4, both at a concentration of 20 ng/mL, was added. The culture medium was changed every two days and supplemented with sufficient cytokines. BMDCs were cultured until day 7.

*2.10. Flow Cytometry.* BMDCs were harvested and washed with 1 mL of phosphate-buffered saline (PBS). The cell suspensions were prepared using conventional methods, and 100 $\mu$L of the fluorescently labeled monoclonal antibodies, CD40-FITC, CD80-PE, CD86-PE, and MHC-II-PE, were added, followed by incubation in the dark for 30 min at room temperature (20–25°C). After centrifugation, 1 mL of PBS buffer solution was aseptically added to wash the cells. Finally, the cells were subjected to flow cytometry (FC500, American Beckman Coulter Company), and the percentages of CD40, CD80, CD86, and MHC-II in BMDCs were analyzed using software.

*2.11. ELISA.* The supernatant of BMDCs was acquired after culture for 7 days using centrifugation (8 min, 2,000 rpm) and frozen at −80°C until detection. The levels of PGE2 (EK8103-01), IL-23 (CSB-E08462r), TNF-$\alpha$ (EK3823-01), and IL-10 (EK3101-01) in BMDCs were detected using ELISA kits, according to the manufacturer's instructions. Absorbance was read at 450 nm.

*2.12. Statistical Analysis.* Data are presented as the mean ± SD. One-way analysis of variance (ANOVA) using SPSS software was used to define the significant differences among the control and the drug groups. A $P < 0.05$ with a 95% confidence interval was considered statistically significant.

## 3. Results

*3.1. The Effects of the Ethyl Acetate Part of EMS on the Clinical Index in AA Rats.* The paw swelling and polyarthritis index were defined to assess the severity of arthritis and the anti-inflammatory effects of medicines. According to Figure 1, the secondary inflammatory reaction appeared around days 14–17, culminating around days 20–23, and the weight of AA rats increased slowly and obviously less compared with the normal rats ($P < 0.01$). On day 28, the ethyl acetate part of EMS treatment increased the weight of rats, and this change was faster compared with the model group. No significant difference in body weight was observed between the EMS and MTX groups during the administration period (Table 1). Compared with the AA model group, the ethyl acetate fraction of the EMS treatment not only ameliorated the effect on the paws of AA rats (Figure 1(a)) and obviously inhibited paw swelling (from 1.0 to 0.7) but also significantly reduced the polyarthritis index (from 12 to 9), $P < 0.01$ (Figures 1(b) and 1(c)). This result suggested that the ethyl acetate part of EMS treatment is effective and safe to use in AA rats.

*3.2. The Effects of the Ethyl Acetate Part of EMS on Ankle Pathology in AA Rats.* Histopathological analysis is considered a great index of clinical signs. To define the effectiveness of the ethyl acetate part of EMS, its antiarthritis effects were analyzed using histological analysis. As shown in Figure 2, the model group expressed severe arthritic symptoms, and synovial tissues infiltrated by many inflammatory cells, accompanied by synovial proliferation, bone erosion, and destruction (compared to the normal group, $P < 0.01$). The results suggested that EMS (3 g/kg) and MTX (0.5 mg/kg) all markedly reduced these histological severity scores (compared to the model group, $P < 0.01$). EMS (0.75 g/kg, 1.5 g/kg) also protected the tissues from histological damage ($P < 0.05$). The results showed that the ethyl acetate part of EMS has a protective effect in AA rats.

*3.3. The Effects of the Ethyl Acetate Part of EMS on the Spleen and Thymus Index in AA Rats.* The spleen and thymus index, a sign of organ changes in the body, accompanied the development of inflammation. As shown in Figure 3, compared to the normal group, model rats, spleen, and thymus index obviously increased ($P < 0.01$). EMS (3 g/kg) and MTX (0.5 mg/kg) significantly reduced the spleen and thymus index, compared to the AA group ($P < 0.01$). EMS at 0.75 and 1.5 g/kg reduced the thymus index, $P < 0.05$ (Figure 3(b)), and at 1.5 g/kg, it significantly reduced the spleen index compared to the AA group ($P < 0.01$; Figure 3(a)). These results suggest that EMS can inhibit the spleen and thymus index in AA rats.

*3.4. The Effects of the Ethyl Acetate Part of EMS on T- and B-Cells Proliferation in AA Rats.* T- and B-cells participate in human-specific immunity. The viability of T- and B-cells was determined using the CCK-8 assay (bestbio biological). As shown in Figure 4, compared to the normal group, both T- and B-cell proliferation increased in the model group. Furthermore, EMS (1.5 g/kg, 3 g/kg) and MTX (0.5 mg/kg) reduced Con A-induced T-lymphocyte proliferation and LPS-induced B-lymphocyte proliferation (compared to the model group, $P < 0.01$). In addition, EMS (0.75 g/kg) effectively suppressed T-lymphocyte proliferation (compared to the model group, $P < 0.05$) (Figure 4(a)). Additionally, the ethyl acetate part of EMS should regulate T- and B-cell function by inhibiting their viability.

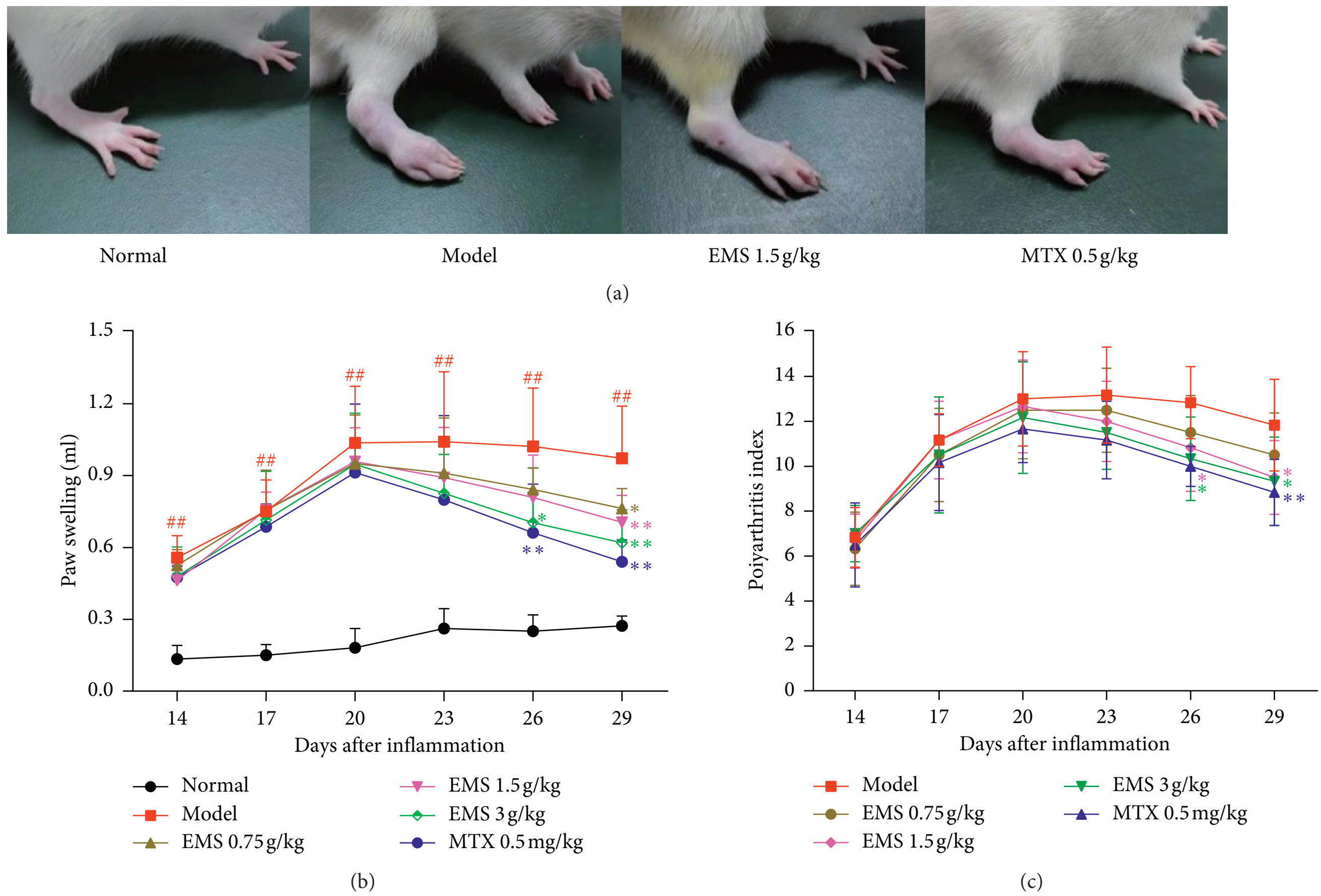

Figure 1: The effects of the ethyl acetate part of EMS on the clinical index in AA rats. (a) The effects of the ethyl acetate part of EMS on the paw in AA rats (day 29). The effects of the ethyl acetate part of EMS on paw swelling (b) and polyarthritis index (c) in AA rats. $^{##}P < 0.01$ versus the normal group, $^{*}P < 0.05$, $^{**}P < 0.01$ versus the model group ($n = 6$).

Table 1: The effects of the ethyl acetate part of EMS on the change of rat weight in AA rats.

| Groups | Dose (mg/kg) | *d*0 | *d*7 | *d*14 | *d*21 | *d*28 |
|---|---|---|---|---|---|---|
| Normal | — | 162 ± 2.95 | 218 ± 6.74 | 276 ± 18.97 | 308 ± 12.48 | 337 ± 19.00 |
| Model | — | 163 ± 2.66 | 215 ± 10.81 | 256 ± 19.71$^{#}$ | 272 ± 12.24$^{##}$ | 280 ± 22.59$^{##}$ |
| EMS | 0.75 g/kg | 163 ± 3.50 | 222 ± 8.86 | 258 ± 19.80 | 291 ± 28.41 | 300 ± 19.07$^{*}$ |
| EMS | 1.5 g/kg | 162 ± 3.08 | 211 ± 5.73 | 260 ± 6.66 | 286 ± 17.95 | 310 ± 26.72 |
| EMS | 3 g/kg | 163 ± 4.65 | 219 ± 10.35 | 263 ± 7.80 | 298 ± 23.33 | 309 ± 28.20 |
| MTX | 0.5 mg/kg | 162 ± 2.66 | 223 ± 4.58 | 258 ± 18.01 | 303 ± 27.77 | 309 ± 24.62 |

$^{##}P < 0.01$ versus the normal group, $^{*}P < 0.05$ versus the model group ($n = 6$).

*3.5. The Effects of the Ethyl Acetate Part of EMS on the Expression of Surface Costimulatory Molecules in BMDCs.* It is well known that DCs are specialized in antigen presenting, which plays an indispensable role in the onset of autoimmune diseases. To determine the effects of EMS on BMDC maturation and activation, CD40, CD80, CD86, and MHC-II were detected on DCs using flow cytometry. The morphology of BMDCs was observed using an inverted microscope (motic, AE2000LED) (Figure 5(a)). As shown in Figure 5, flow cytometry results showed that the expression of CD40, CD80, CD86, and MHC-II were upregulated in the model group (compared to the normal group, $P < 0.01$). EMS (3 g/kg) treatment clearly suppressed the expression of CD80, CD86, and MHC-II ($P < 0.01$). In addition, EMS (0.75 g/kg) decreased the expression of CD86 compared with AA rats ($P < 0.01$). Moreover, EMS (1.5 g/kg) and MTX (0.5 mg/kg) suppressed the expression of CD40, CD80, CD86, and MHC-II in BMDCs compared with AA rats. The ethyl acetate part of EMS treatment inhibited the effect on BMDCs in AA rats in a dose-dependent manner. These results suggest that EMS inhibits BMDC maturation in AA rats.

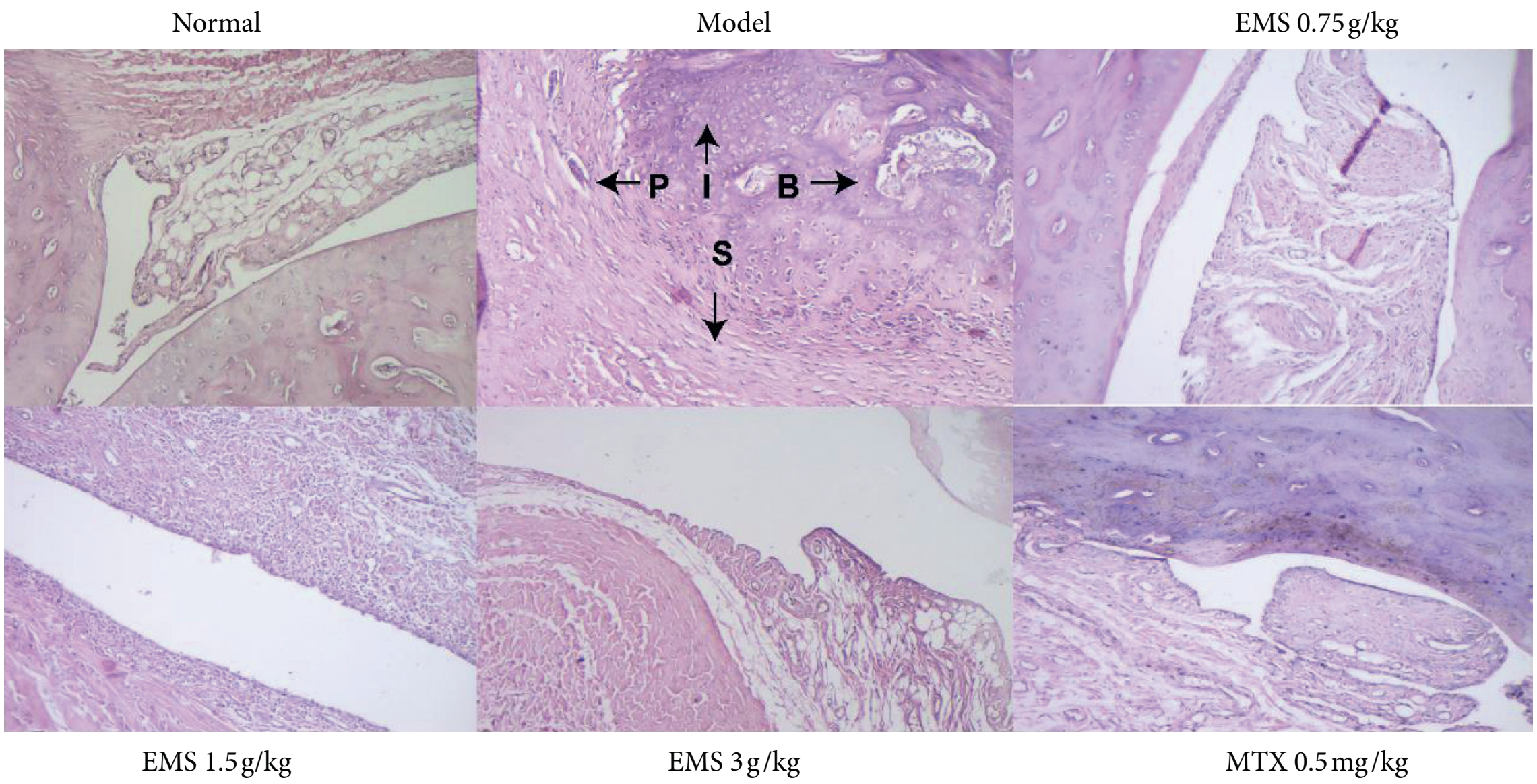

(a)

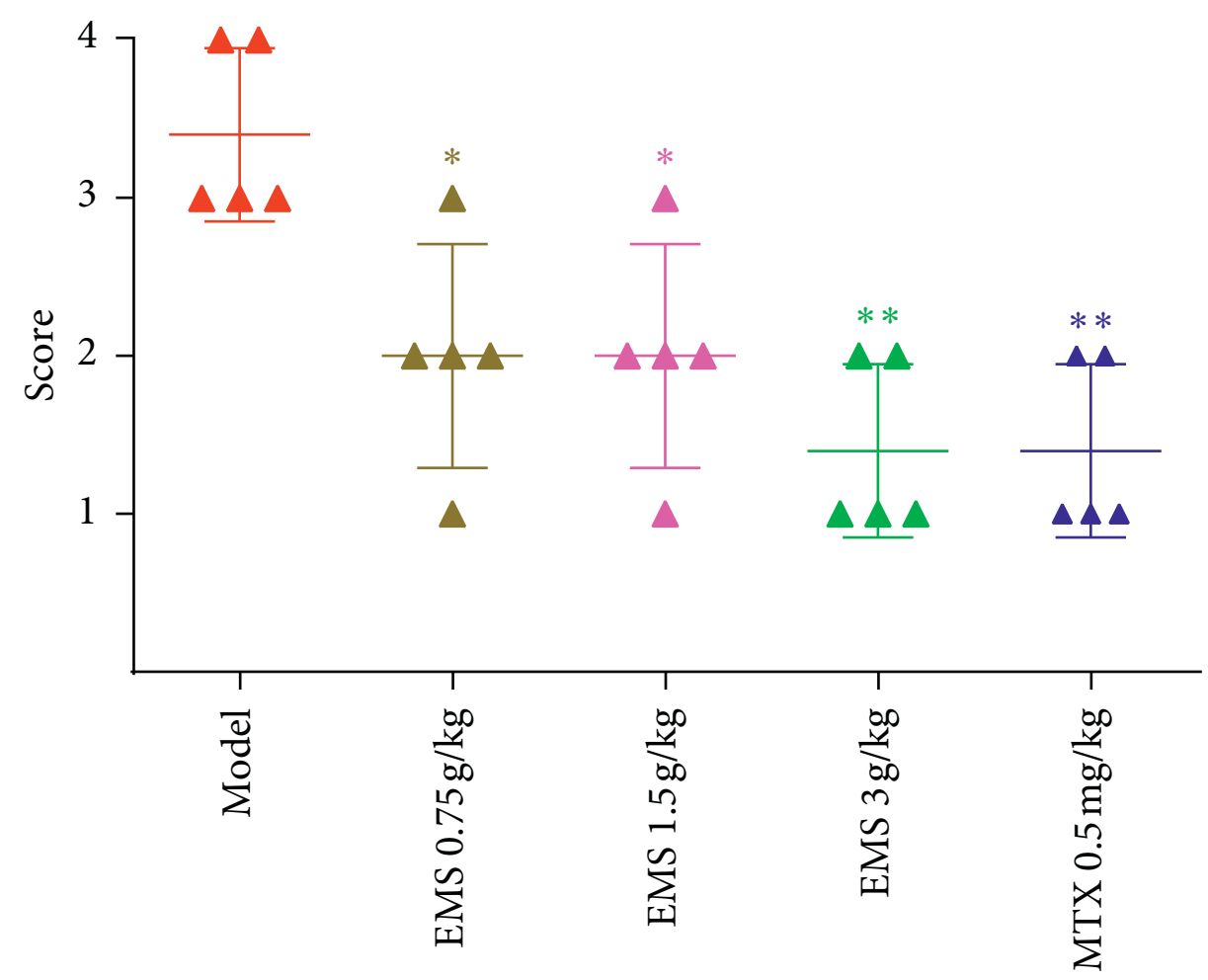

(b)

Figure 2: The ethyl acetate part of EMS effects on ankle pathology in AA rats was detected using HE staining. (a) Photographs of representative EMS effects on the ankle pathology in AA rats (HE, ×100 magnification). P: pannus, B: bone erosion, S: synovial proliferation, and I: inflammatory cells infiltration. (b) The scores of histopathology in the rats were calculated. $^{*}P < 0.05$, $^{**}P < 0.01$ versus the model group ($n = 5$).

*3.6. The Effects of the Ethyl Acetate Part of EMS on BMDCs Supernatant Cytokines in AA Rats.* With the development of RA, intracellular inflammatory factors have changed. To evaluate the ability of the ethyl acetate part of EMS to modulate cytokine production in BMDCs, proinflammatory cytokines (TNF-$\alpha$ and IL-23), inflammatory factors (PGE2), and anti-inflammatory cytokine (IL-10) were detected in the supernatant of BMDCs. As shown in Figure 6, compared with the normal group, the concentration of TNF-$\alpha$, PGE2, and IL-23 in the supernatant of BMDCs increased in the model group ($P < 0.01$). By contrast, the concentration of IL-10 dramatically declined in the model group ($P < 0.01$) (Figure 6(d)). EMS (0.75, 1.5, and 3 g/kg) or MTX (0.5 mg/kg) significantly inhibited the concentration of proinflammatory cytokines (TNF-$\alpha$ and IL-23) and inflammatory factors (PGE2) (compared to the model group, $P < 0.01$). Moreover, the levels of anti-inflammatory cytokine (IL-10) increased in a dose-dependent manner after EMS treatment ($P < 0.01$). These results suggest that EMS inhibits BMDC function to secrete inflammatory factors.

## 4. Discussion

Previous research has shown that EMS water extract has a therapeutic effect on AA rats, significantly reducing the pathological index and paw swelling in rats with arthritis [12]. In order to further study the effective antiarthritis parts of EMS, these were extracted using different organic solvents, such as dichloromethane, ethyl acetate, chloroform, and petroleum ether. Dichloromethane and ethyl acetate were the two parts of EMS treatment that had an effective effect on AA rats [8]. Through a comprehensive evaluation of the arthritic indicators, the ethyl acetate part of EMS might be an effective anti-inflammatory part. In this study, an aqueous solution of EMS was extracted with petroleum ether, which was directly extracted with ethyl acetate. Therefore, we obtained the ethyl acetate part of EMS. The ethyl acetate part of EMS (0.75, 1.5, and 3 g/kg) was administered to AA rats. The results showed that the ethyl acetate part of the EMS treatment obviously reduced paw swelling (Figure 1(b)) and the polyarthritis index (Figure 1(c)), improving the histological severity scores (Figure 2), which is consistent with our previous study [8]. However, the results provided the basis of pharmacodynamics and inspired us to continue exploring the mechanism of ethyl acetate in EMS antiarthritis therapy. In addition, berberine and atractylodin were detected from the EMS ethyl acetate part using HPLC in a previous study [8]. Berberine has a positive effect in AA rats, reducing the concentration of proinflammatory cytokines (IL-1$\beta$, IL-6, and IL-17) and increasing the concentration of anti-inflammatory cytokines (IL-10 and TGF-$\beta$) [13, 14]. Furthermore, berberine and atractylodin could protect collagen-induced arthritis rats by regulating the DC function by suppressing maturation or promoting apoptosis [15, 16]. Therefore, we speculate that the antiarthritis effect of EMS may be due to the regulation of DC maturation.

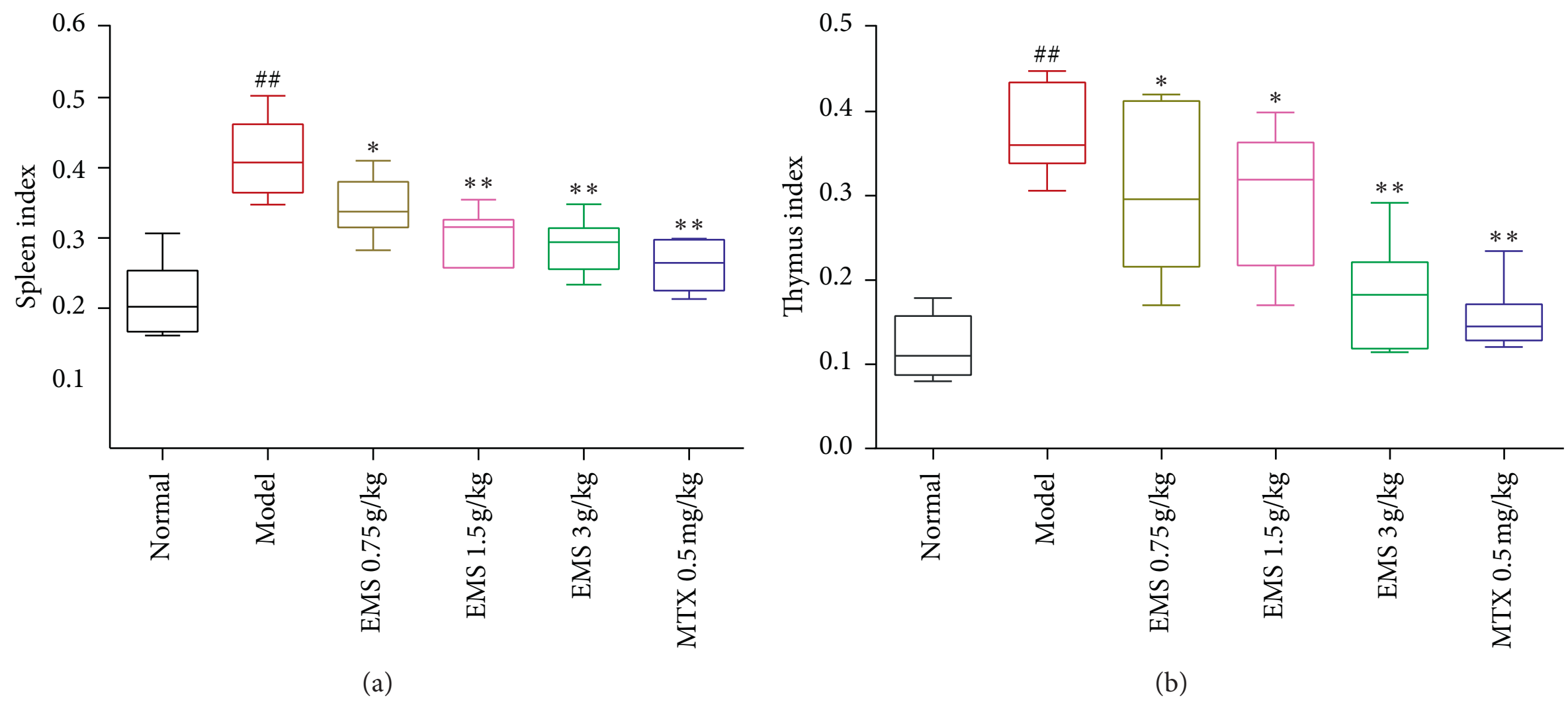

FIGURE 3: The effects of the ethyl acetate part of EMS on spleen and thymus index in AA rats. (a) Spleen index; (b) thymus index; $^{\#\#}P < 0.01$ versus the normal group. $^{*}P < 0.05$, $^{**}P < 0.01$ versus the model group ($n = 6$).

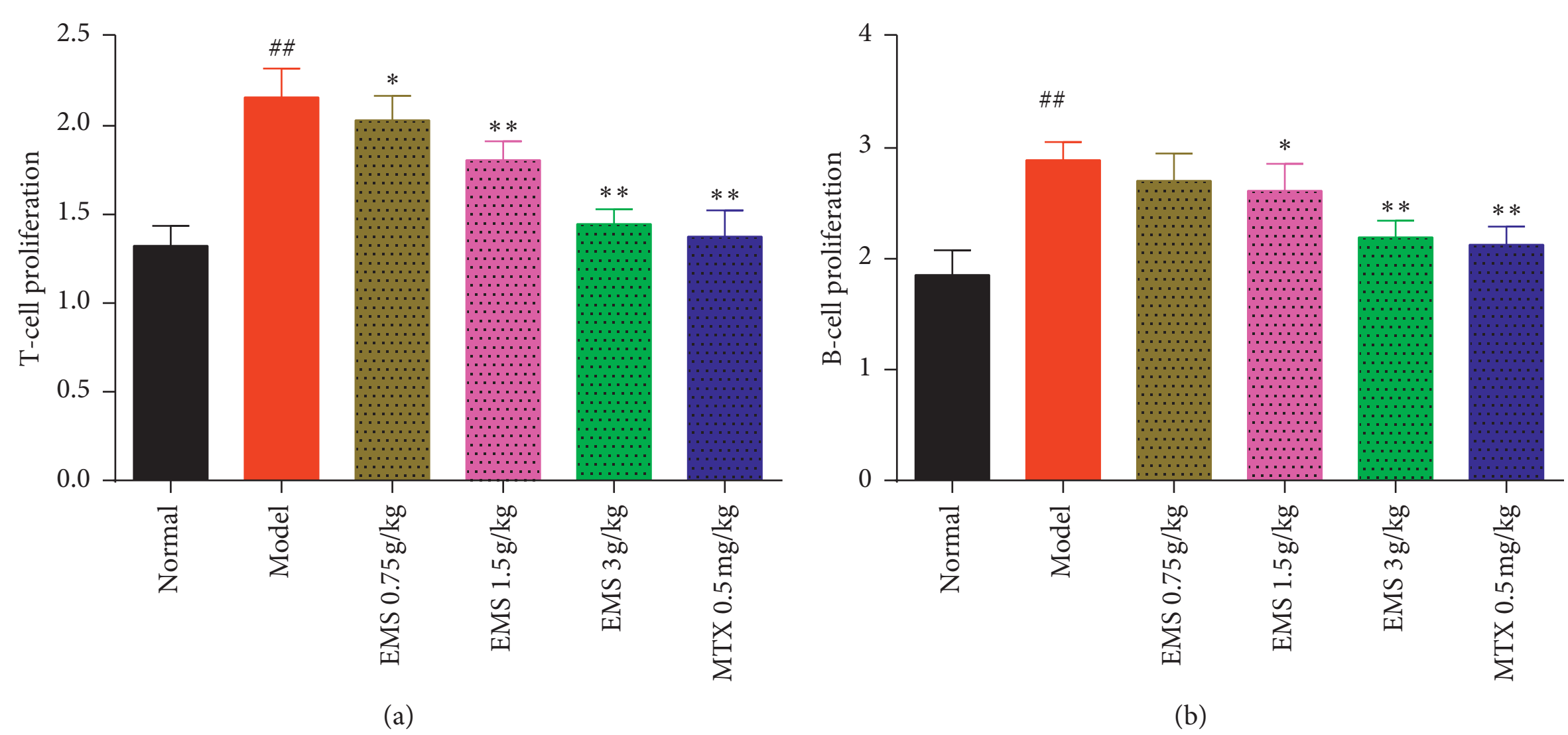

FIGURE 4: The effects of the ethyl acetate part of EMS on T- and B-cell proliferation in AA rats. (a) T-cell proliferation; (b) B-cell proliferation; $^{\#\#}P < 0.01$ versus the normal group. $^{*}P < 0.05$, $^{**}P < 0.01$ versus the model group ($n = 6$).

We found that the ethyl acetate part of EMS obviously reduced the spleen index and thymus index (Figure 3) and inhibited the proliferation of T- and B-cells (Figure 4). It has been reported that cytotoxic T-lymphocyte-associated antigen-4 (CTLA-4) blocks the interaction between APC and T-cells, in which the mode of action is integrated with CD80 and CD86 receptors on APC, preventing CD28 on the surface of T-lymphocytes [17]. This protein receptor has been found to play a crucial role in autoimmune disease by blocking the interplay between B-cells, APC, and T-cells. It was suggested that the fact that the ethyl acetate part of EMS alleviated arthritis may be related to the inhibition of T- and B-cells. As it has been reported, DCs are specialized in antigen presenting, which plays an indispensable role on the onset of autoimmune diseases. Mature DCs highly express surface stimulating molecules, such as CD40, CD80, CD86, and MHC-II. As an important indicator to identify DCs, flow cytometry can accurately and quickly detect the expression of CD40, CD80, CD86, and MHC-II on their surface. To determine the effects of EMS on BMDC maturation, CD40, CD80, CD86, and MHC-II were detected on DCs using flow cytometry. The results showed that the expression of CD40, CD80, CD86, and MHC-II was upregulated in AA rats. Treatment with EMS clearly suppressed the expression of CD40, CD80, CD86, and MHC-II in BMDCs compared with AA rats. Therefore, the ethyl acetate part of the EMS antiarthritis mechanism may be related to the inhibition of BMDC maturation.

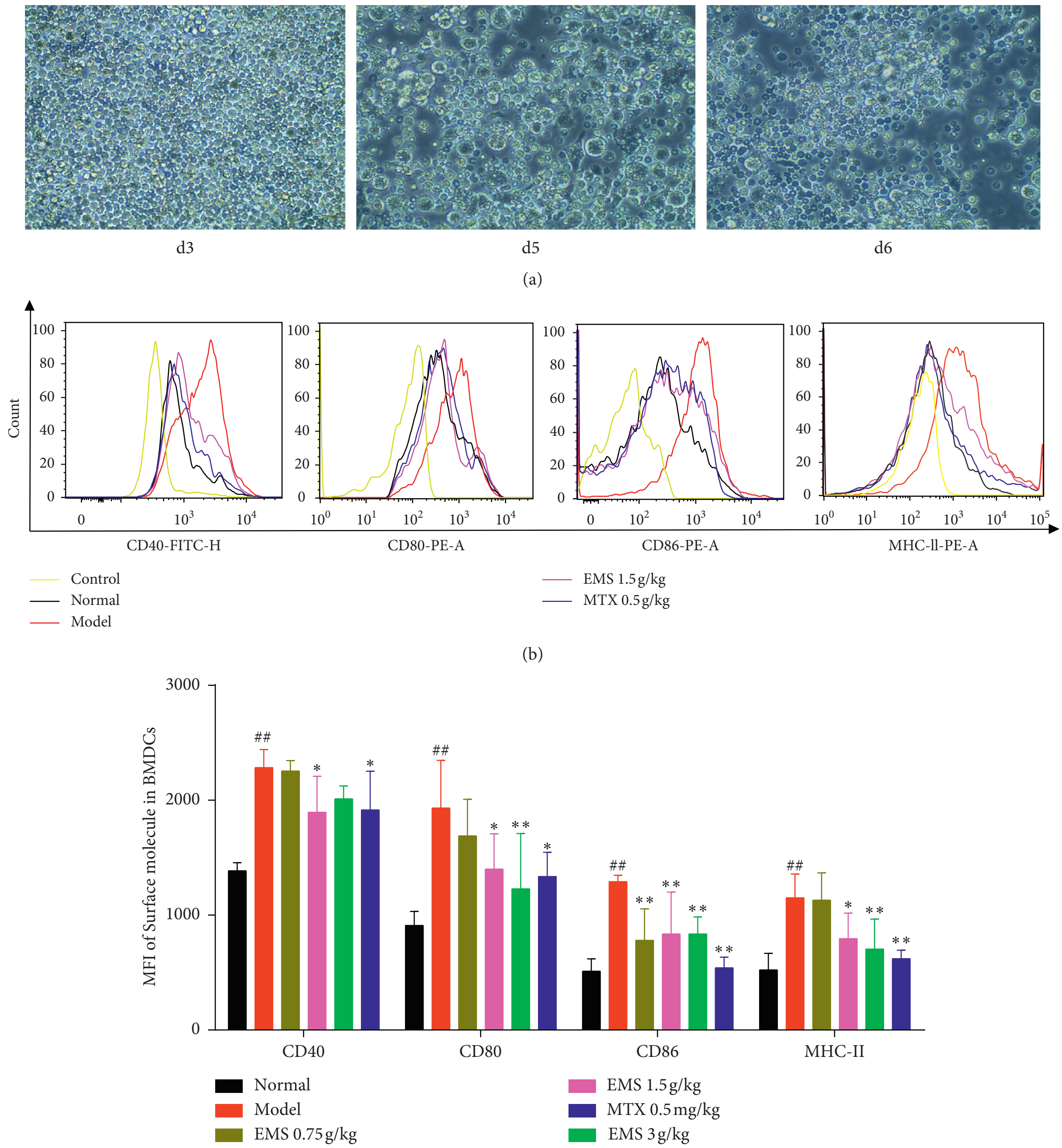

FIGURE 5: The effects of the ethyl acetate part of EMS on the expression of CD40, CD80, CD86, and MHC-II on BMDCs. (a) The morphology of bone marrow-derived dendritic cells (x200 magnification). (b) CD40, CD80, CD86, and MHC-II were detected by flow cytometry. (c) The MFI of CD40, CD80, CD86, and MHC-II in BMDCs were calculated. $^{##}P < 0.01$ versus the normal group. $^{*}P < 0.05$, $^{**}P < 0.01$ versus the model group ($n = 4$).

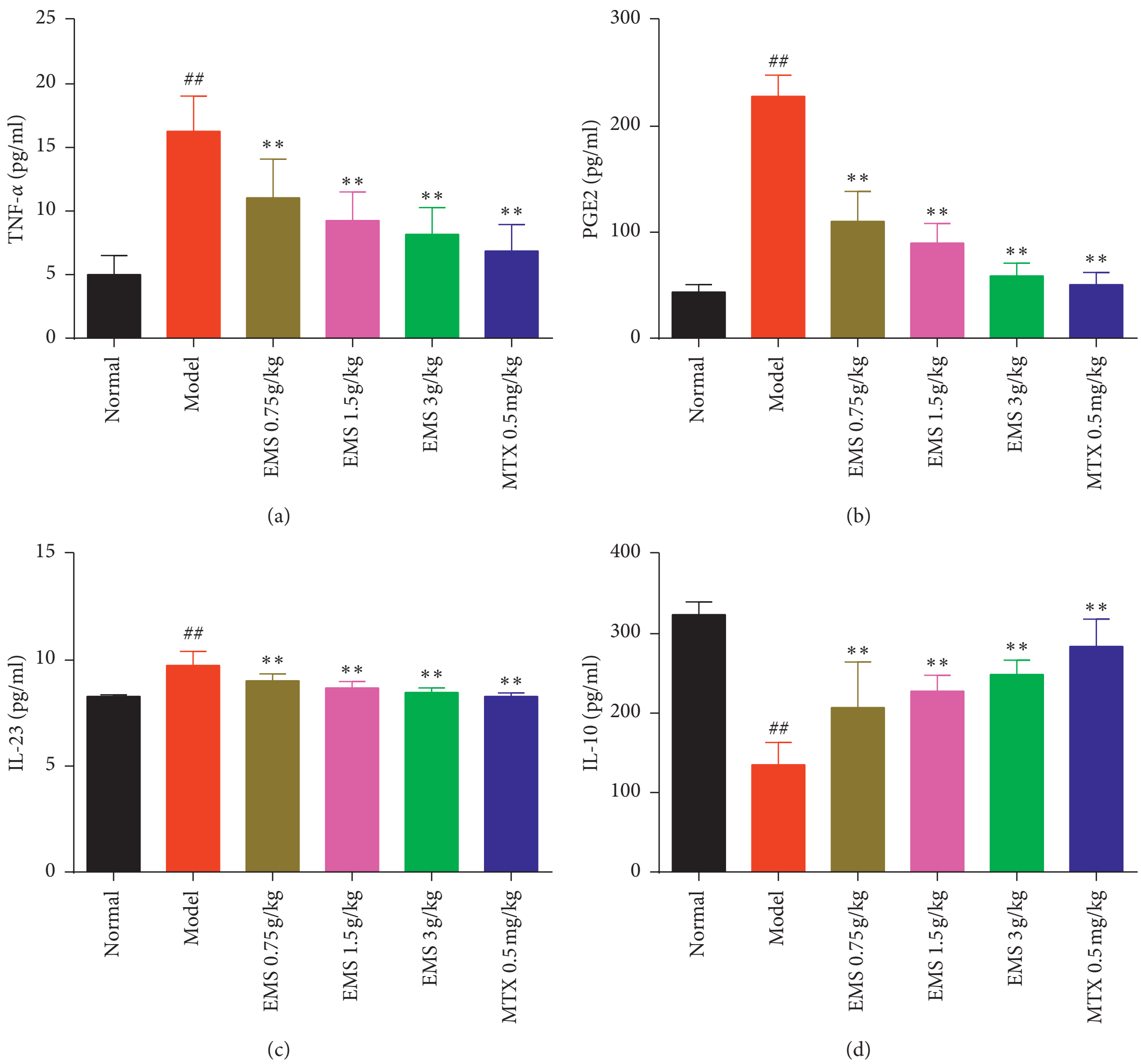

Figure 6: The effects of the ethyl acetate part of EMS on cytokines production in BMDCs of AA rats. The levels of TNF-$\alpha$, PGE2, IL-23, and IL-10 in BMDCs were measured by ELISA. (a) TNF-$\alpha$, (b) PGE2, (c) IL-23, and (d) IL-10, $^{\#\#}P < 0.01$, versus the normal group. $^{**}P < 0.01$ versus the model group ($n = 6$).

IL-6 blocked tocilizumab and the IL-1 blocker anakinra have also been used to treat RA [18, 19]. DCs induce T-cell differentiation into Th1 and Th17 cells, which secrete IL-12 and IL-23 to exacerbate RA disease [20, 21]. IL-10 plays a crucial role in regulating Treg cell differentiation and enhancing the effect of Treg cell inhibition of TH17 cells to mitigate the degree of joint inflammation in AA mice [22, 23]. Meanwhile, we detected that the ethyl acetate part of EMS effectively reduced the concentration of proinflammatory cytokines TNF-$\alpha$ (Figure 6(a)) and IL-23 (Figure 6(c)) and increased the level of anti-inflammatory cytokine IL-10 (Figure 6(d)). PGE2 is an important metabolite of arachidonic acid and plays an important regulatory role in the immune response. PGE2 was detectable in the synovium of patients with arthralgia [24]. It has been reported that PGE2 participates in the expression of prostaglandin E4 and nuclear factor kappa-B in DCs, which upregulate the level of cAMP protein by inducing IL-23 Subunit p19 in DCs [25, 26]. We found that the concentration of PGE2 in the supernatant of BMDCs increased in AA rats. EMS treatment significantly inhibited the concentration of PGE2 (Figure 6(b)). Therefore, the mechanism of alleviation of inflammation by inhibiting the expression of BMDCs may be related to a variety of signaling pathways. This might provide evidence to study the mechanism of the active pathway of DCs in RA. In future research, we will focus on the mechanism of the specific action of EMS treatment on inhibiting the function of DC and its relationship with the participation of molecular signals.

## 5. Conclusions

These results suggest that the ethyl acetate part of EMS has better protective effects on AA rats, probably by regulating the maturation of BMDCs and modulating the balance of cytokines.

## Data Availability

The data used to support the findings of this study are available from the corresponding author upon request.

## Conflicts of Interest

The authors have no conflicts of interest to declare.

## Authors' Contributions

XYJ and NW designed and performed experiments and revised the manuscript. JMD designed and performed experiments and wrote the manuscript. ZHX, ML, and MLL performed experiments. All authors read and approved the manuscript.

## Acknowledgments

This work was supported by the National Natural Science Foundation of China (81603362) and by the Anhui Natural Science Foundation of China (1708085QC77).